\documentclass[submission,copyright,creativecommons]{eptcs}
\providecommand{\event}{ICLP 2019} 

\usepackage{graphicx}
\usepackage{mathptmx}
\usepackage{tikz-qtree}
\usepackage{pgfplots}
\usepackage{filecontents}
\usepackage{csvsimple}
\usepackage{amsfonts,amsmath}
\usepackage{url}
\usepackage{verbatim}
\usepackage{color}

\definecolor{lgray}{gray}{0.95}
\definecolor{lblue}{rgb}{0.90,0.90,1.00}
\definecolor{lyellow}{rgb}{1.00,1.00,0.70}

\usepackage{listings}
\newtheorem{ex}{Example}

\newenvironment{codex}{\small\verbatim}{\endverbatim\normalsize}

\lstnewenvironment{code}
    {\lstset{}%
      \csname lst@SetFirstLabel\endcsname}
    {\csname lst@SaveFirstLabel\endcsname}
\newcommand{\BI}[0]{\begin{itemize}}
\newcommand{\EI}[0]{\end{itemize}}
\newcommand{\I}[0]{\item}
\newcommand{\BE}[0]{\begin{enumerate}}
\newcommand{\EE}[0]{\end{enumerate}}

\newcommand{\BX}[0]{\begin{ex}}
\newcommand{\EX}[0]{\end{ex}}
\newcommand{\BV}[0]{\begin{verbatim}}
\newcommand{\EV}[0]{\end{verbatim}}

\newcommand{\BF}[0]{\begin{filecontents*}{data.csv}}

\newcommand{\BQ}[0]{\color{blue}\begin{quote}}
\newcommand{\EQ}[0]{\end{quote}\color{black}}

\def \bscale1 {0.25}
\def \bscale {0.25}

\usepackage{hyperref}
\usepackage{booktabs}

\begin{document}

\title{Lazy Stream Programming in Prolog}

\author{
          Paul Tarau\email{
          paul.tarau@unt.edu}
          \and
          Jan Wielemaker\email{
          J.Wielemaker@vu.nl}
          \and
          Tom Schrijvers\email{
          tom.schrijvers@cs.kuleuven.be}
}

\def \authorrunning{Paul Tarau, Jan Wielemaker and Tom Schrijvers}
\def\titlerunning{Lazy Stream Programming in Prolog}

\maketitle

\begin{abstract}
In recent years, stream processing has become a prominent approach for incrementally handling large amounts of data, with special support and libraries in many programming languages. Unfortunately, support in Prolog has so far been lacking and most existing approaches are ad-hoc. To remedy this situation, we present {\em lazy stream generators} as a unified Prolog interface for stateful computations on both finite and infinite sequences of data that are produced incrementally through I/O and/or algorithmically.

We expose stream generators to the application programmer in two ways: 1) through an abstract sequence manipulation API, convenient for defining custom generators, and 2) as idiomatic lazy lists, compatible with many existing list predicates.  We define an algebra of stream generator operations that extends Prolog via an embedded language interpreter, provides a compact notation for composing generators and supports moving between the two isomorphic representations.

As a special instance, we introduce answer stream generators that encapsulate the work of coroutining first-class logic engines and support interoperation between forward recursive {\em AND-streams} and backtracking-generated {\em OR-streams}. 


\end{abstract}

\hrule

\section{Introduction}

Initial design as well as evolution of successful programming languages often walks a fine line between semantic purity and pragmatic expressiveness. With its declarative roots and creative pragmatic additions Prolog is a long-time survivor in the complex ecosystem of programming languages. We believe that its longevity is due not only to its elegant semantics but also to its creative adaptations to emerging programming language features that respond to evolving software development requirements.

\emph{Stream processing}---now prevalent in widely used programming languages
languages like Java, Python, C\#, go or JavaScript---offers a uniform and
(mostly) declarative view on processing finite and infinite\footnote{We use
``infinite'' here as a short hand for data or computation streams of
unpredictable, large or very large size.} sequences. Besides the expressiveness
boost it provides, its advent has been driven by the need for processing big
data. This big data problem manifests itself in static incarnations like very
large training sets for machine learning, or as dynamic event streams coming
from Web search queries and clicks, or from sensor networks supporting today's
fast spreading IoT infrastructure.

The main goal of this paper is to extend Prolog with state-of-the-art lazy
stream processing capabilities like those available in other languages.
While some languages facilitate such an extension with features like
generalized iterators (Python) or a lazy evaluation semantics (Haskell), 
Prolog presents two major obstacles that make this task particularly challenging.

The first obstacle is presented by Prolog's fixed depth-first search resolution
and strict evaluation semantics. While Prolog's depth-first search mechanism
can be complemented with alternative search strategies, as shown in \cite{tor}
by overriding its disjunction operator, the evaluation mechanism remains
ultimately eager. When programming with lists or DCGs, one chains recursive
steps in the body of clauses connected by conjunctions. 

The second obstacle, a consequence of Prolog's incremental evolution as a
programming language, is the presence of procedural state-management and I/O
constructs that are interleaved with its native declarative programming
constructs. These range from random generator streams to file and socket I/O
and dynamic database operations.  While monadic constructs in functional
languages \cite{moggi:monads,wadler93:cont}  offer  a unified view of
declarative and procedural state-management operations, most logic programming
languages still lack a unified approach providing a uniform interface to this
mix of declarative and procedural language constructs.

We manage to overcome these obstacles and provide lazy stream
processing for Prolog in a way that uniformly encapsulates
different streaming mechanisms---state transformers, lazy lists and first-class logic engines 
\cite{tarau:cl2000,padl09inter,bp2011}, 
recently added to SWI-Prolog\footnote{
\url{http://www.swi-prolog.org/pldoc/man?section=engines}}, into a set of operations organized compositionally in the form of {\em 
stream generators}.
Our generators work in a way similar to Python's {\em yield} mechanism \cite{pyref,beazley09} and 
they share features with coroutining constructs now present in a several  programming languages including C\#, go, Javascript and Lua. At the same time, 
they lift Prolog's expressiveness with lazy evaluation mechanisms similar to non-strict
functional programming languages like Haskell \cite{hudak07} or functional-logic languages like Curry \cite{antoy05}.

We organize our generators as an algebra, wrapped as a library module with a declarative interface, to avoid exposing  operations requiring an implementation with a clear procedural flavor to the Prolog application programmer.

By defining a functor that transports operations between isomorphic
generators and lazy lists, we offer a choice between 
abstract sequence operations and the concrete list view familiar to Prolog users.


The main contributions of this paper are:
\BI
\I We present a simple and clean approach for setting up lazy streams, which
   uniformly encapsulates algorithms, lists, first-class logic engines and other data sources.
\I We show how to expose lazy streams in the form of lazy Prolog lists
   that, just like conventional lists, can be inspected and decomposed with unification.
   Under the hood, lazy lists use attributed variables and destructive updates to
   extend the list when needed.
\I We have implemented our approach in several libraries:
   \BE 
   \I Our \texttt{lazy\_streams} library
features a dozen generator predicates (stream sources), an API to query
them, a set of generator  operations, a generator expression interpreter
offering a declarative view of these operations and an interface to
the next library. 
   \I Our \texttt{lazy\_lists} library provides a dozen generator predicates for 
      directly setting up lazy lists.
   \I Our \texttt{pure\_input} library provides a range of predicates for reading
      files and sockets backed by lazy lists.
   \EE
   The first is available as an SWI-Prolog
library package\footnote{%
\url{https://github.com/ptarau/AnswerStreamGenerators/raw/master/lazy_streams-0.5.0.zip}},
  while the other two are bundled as a SWI-Prolog standard libraries\footnote{
\url{http://www.swi-prolog.org/pldoc/doc/_SWI_/library/lazy_lists.pl}}
\footnote{
\url{http://www.swi-prolog.org/pldoc/doc/_SWI_/library/pure_input.pl}}.
\EI

The rest of the paper is organized as follows.
Section \ref{ov} demonstrates our approach with some examples.
Section \ref{impl} describes implementation of lazy stream generator constructors and 
their  interface.
Section \ref{alg} introduces several operations on generators and overviews the
embedded language interpreter organizing them as an algebra of generator combinators.
Section \ref{other} describes lazy functional language style generator operations
and an example of I/O stream generator.
Section \ref{ll} overviews implementation of lazy lists using attributed variables
and introduces the iso-functor connecting them to  lazy stream generators.
Section \ref{disc} compares and discusses alternative implementation options of lazy list and stream generators.
Section \ref{rel} overviews related work and
section \ref{conc} concludes the paper. 

\section{Overview}\label{ov}

This section briefly introduces our lazy streams  with a few examples.

The generators {\tt pos/1} and {\tt neg/1} produce the infinite streams of positive
and negative integers. With {\tt map/4} we combine these two streams element-wise;
here they annihilate each other with {\tt plus/3}. With {\tt show/2} we display
the first 10 elements of the resulting constant stream of zeroes.
\begin{codex}
?- pos(P),neg(N),map(plus,P,N,Zero),show(10,Zero).
[0, 0, 0, 0, 0, 0, 0, 0, 0, 0].
\end{codex}
With \texttt{list/2} we turn the regular list \texttt{[a,b,c]} into a stream.
Then {\tt convolution/3} computes its Cartesian product with the positive
integers, following a convolution approach, and {\tt show/2} displays the first 16 elements. 
\begin{codex}
?- pos(P),list([a,b,c],L),convolution(P,L,C),show(16,C).
[1-a, 1-b, 2-a, 1-c, 2-b, 3-a, 2-c, 3-b, 4-a, 3-c, 4-b, 5-a, 4-c, 5-b, 6-a, 5-c].
\end{codex}
We also provide an embedded language interpreter to concisely 
express algebraic operations on streams. Here we use it to create
the Cartesian product of the stream \texttt{[a,b]} with the stream
\texttt{[1,2,3]}, the latter abbreviated by the shorthand \texttt{(1:4)}.
The \texttt{in\_/2} predicate enumerates the elements of the resulting stream
through backtracking, like \texttt{member/2} does for regular lists.
\begin{codex}
?- X in_ [a,b]*(1:4).
X = a-1 ; X = b-1 ; X = b-2 ; X = a-2 ; X = b-3 ; X = a-3 .
\end{codex}

\section{Implementing Lazy Stream Generators}\label{impl}
  
Generators are created by a family of constructors, encapsulating sequences
produced algorithmically or as a result of state transformers interfacing
Prolog with the ``outside world'', a design philosophy similar to that of
monads in functional languages.

\subsection{The Stream Generator Interface}

A generator is represented by a \emph{closure} (assumed deterministic), which is a term that can be
called with an additional argument to yield the next element in the stream. For
instance, the generator for the constant stream of \texttt{1}s is
\texttt{=(1)}, where \texttt{call(=(1),X)} instantiates \texttt{X} to the next
element, which is always \texttt{1}.  Typically the other arguments of the term
represent the state and parameters of the generator, like \texttt{1} in
\texttt{=(1)}. We require that the closure has at least one argument, which
never takes the reserved value \texttt{done} in the course of its operation.
When the closure has no more elements to yield, it fails.

Our {\tt ask/2} predicate provides a basic interface to interact with generators:\\
\begin{tabular}{ccc}
\begin{minipage}{0.4\textwidth}
\begin{code}
ask(E,_):-is_done(E),!,fail.
ask(E,R):-call(E,X),!,R=X.
ask(E,_):-stop(E),fail.
\end{code}
\end{minipage}
&
where
&
\begin{minipage}{0.4\textwidth}
\begin{code}
is_done(E):-arg(1,E,done).
stop(E):-nb_setarg(1,E,done).
\end{code}
\end{minipage}
\end{tabular}

This code {\tt call}s the generator to produce an element. The first time the
generator fails, \texttt{stop/1} writes \texttt{done} into its first argument
in a non-backtrackable fashion (and then propagates the failure). Subsequent 
{\tt ask}s simply read the argument with \texttt{is\_done/1} and never invoke
the generator again. This means that its resources can be garbage collected.

The {\tt in/2} predicate uses {\tt ask/2} to produce the elements on backtracking.
\begin{code}
:-op(800,xfx,(in)).

X in Gen:-ask(Gen,A),select_from(Gen,A,X).

select_from(_,X,X).
select_from(Gen,_,X):-X in Gen.
\end{code}

\paragraph{Basic Stream Generators}
We package basic stream generators into a predicate that sets them up from
given parameters.
For instance, the constant stream is created by the {\tt const/2} predicate
which takes the constant value {\tt C} as an input.
\begin{code}
const(C,=(C)).
\end{code}

The {\tt rand/1} predicate produces the {\tt random/1} stream generator, which relies on externally maintained state
to yield floating point numbers between 0 and 1.
\begin{code}
rand(random).
\end{code}

We can also generate a stream by by incrementally evolving a state:
\begin{code}
gen_next(F,State,X):-arg(1,State,X),call(F,X,Y),nb_setarg(1,State,Y).
\end{code}
Here {\tt State} acts as a container for destructively updated values using the
{\tt nb\_setarg/3} built-in.\footnote{\url{http://www.swi-prolog.org/pldoc/doc_for?object=nb_setarg/3}}


For instance, we can define the stream of natural numbers as an evolving state:
\begin{code}
nat(gen_next(succ,state(0))).
\end{code}

The more general {\tt gen\_nextval/3} predicate supports generators for which
the evolving state does not coincide with the elements of the stream.
\begin{code}
gen_nextval(Advancer,State,Yield):-
  arg(1,State,X1),
  call(Advancer,X1,X2,Yield),
  nb_setarg(1,State,X2).
\end{code}

For instance, this approach is useful to turn a list into a stream.
\begin{code}
list(Xs, gen_nextval(list_step,state(Xs))).

list_step([X|Xs],Xs,X).
\end{code}

%

We have built similar stream generators in the library package 
{\tt lazy\_streams}, for a {\em range} of numbers, turning a finite list into
a an infinite cycle of its elements, as well as stream 
transformers excising a finite slice of a larger, possibly infinite stream,
with taking or dropping an initial segment of a stream as special cases.

\subsection{Answer Stream Generators}

We can encapsulate first class logic engines as generators  when more
complex computations are needed for generating the streams, that  cannot be
expressed as simple step-by-step state transformations.

\subsubsection{SWI Prolog's First-Class Logic Engine Implementation}

A first-class logic engine \cite{tarau:cl2000,bp2011} can be seen as a Prolog
virtual machine that has its own stacks and machine state.  
In their SWI-Prolog
implementation, unlike normal Prolog threads \cite{swi,swi_threads},
they are
not associated with an operating system thread. Instead, one asks an engine for
a next answer with the predicate {\tt engine\_next/2}. Asking an engine for the
next answer attaches the engine to the calling operating system thread and
causes it to run until the engine calls {\tt engine\_yield/1} or its associated
goal completes with an answer, failure or an exception. After the engine yields
or completes, it is detached from the operating system thread and the answer
term is made available to the calling thread. Communicating with an engine is
similar to communicating with a Prolog system through the terminal: the client
decides how many answers it wants returned and what to do with them.

Engines are created with the built-in {\tt engine\_create/3}, that uses a goal
and answer template as input and returns an engine handle as output.
SWI-Prolog's engines are created with minimal dynamic stack space and are
garbage collected when unreachable.

Note that implementing the engine API does not need  a Prolog system that
supports multi-threading. It only assumes that the virtual machine is fully
re-entrant, that it can be queried and that it can stop, yield data and resume
execution as a {\em coroutine}.

\subsubsection{Answer Stream Generators}

The predicate {\tt eng/3} creates a generator as a wrapper for the {\tt
engine\_next(Engine,Answer)} built-in, encapsulating the answers of that engine
as a stream.
\begin{code}
eng(X,Goal,engine_next(Engine)):-engine_create(X,Goal,Engine). 
\end{code}
An alternative constructor, {\tt ceng/3}, is also available if one wants to preserve
the goal and answer template, usable, for instance, to clone
the engine's answer stream, an
operation that makes sense only when the Prolog code it is based on, 
is free of side effects.

\subsubsection{The AND-stream / OR-stream Duality}

A key feature of first-class engines is that they support two ways of producing
a stream of answers: 1) via backtracking (OR-streams), and 2) as part of a
forward moving recursive loop (AND-streams).

The stream generator abstraction makes the user oblivious to this choice of
generation method, and allows us to seamlessly replace one implementation 
for another. Consider for instance the two implementations of the stream of natural
numbers below.
The first implementation generates an AND-stream yielding the elements in a
recursive loop.
\begin{code}
and_nat_stream(Gen):-eng(_,nat_goal(0),Gen).

nat_goal(N):-succ(N,SN),engine_yield(N),nat_goal(SN).
\end{code}

The OR-stream implementation generates the successive elements via backtracking.
\begin{code}
or_nat_stream(Gen):-eng(N, nat_from(0,N), Gen).

nat_from(From,To):- From=To ; succ(From,Next),nat_from(Next,To).
\end{code}

When using engines, 
both AND-streams and OR-streams can be infinite, as in the
case of the generators {\tt or\_nat\_stream} and {\tt and\_nat\_stream}.
While one can see backtracking over an infinite set of answers as
a ``naturally born'' OR-stream, the ability of the engine-based  generators to
yield answers from inside an infinite recursive loop is critical
for generating infinite AND-streams.
Because the choice of generation method is immaterial to the user of the generator,
the implementor can choose the most convenient or efficient approach.

\section{The Generator Algebra}\label{alg}

This section describes a set of stream combinators exposed as an algebra via an
embedded language interpreter.
 
\subsection{Operations on Finite or Infinite Stream Generators}

\paragraph{Sums of Streams}

We define the interleaving of two streams to be their sum. This operation is
captured in the predicate {\tt sum(+Gen1,+Gen2, -NewGen)}, which advances by
asking each generator, in turn, for an answer. When one generator terminates,
it keeps progressing in the other.
\begin{code}
sum(E1,E2,sum_next(state(E1,E2))).

sum_next(State,X):-State=state(E1,E2),ask(E1,X),!,
  nb_setarg(1,State,E2),
  nb_setarg(2,State,E1).
sum_next(state(_,E2),X):-ask(E2,X).
\end{code}
For instance,
\begin{codex}
?- pos(N),neg(M),sum(N,M,E),show(10,E).
[1,-1,2,-2,3,-3,4,-4,5,-5]
\end{codex}
We name this operation the ``sum'' because it is clearly associative and, if the
order of the elements is unimportant (with inputs seen as {\em multisets}), it is also commutative.
 Also, it has
the empty stream, defined such that {\tt ask/2} always fails on it, as its neutral element.

\paragraph{Products of Streams}

The Cartesian product is the product operation on two streams. We can
easily implement it by means of convolution in a first-class logic engine.
Our implementation uses a recursive loop that supports possibly infinite stream
generators by storing their finite initial segments into two lists that start
out empty.
\begin{code}
prod(E1,E2,E):-eng(_,prod_goal(E1,E2),E).

prod_goal(E1,E2):-ask(E1,A),prod_loop(1,A,E1-[],E2-[]).
\end{code}

The algorithm, expressed by the predicate {\tt prod\_loop},
alternates between both generators while neither is done.
After that, it keeps progressing in the remaining generator until 
that too is exhausted. Each time a generator produces a new element,
it is paired with the previously produced elements of the other generator,
which are stored in a list.
\begin{code}
prod_loop(Ord1,A,E1-Xs,E2-Ys):-
  flip(Ord1,Ord2,A,Y,Pair),
  forall(member(Y,Ys),engine_yield(Pair)),
  ask(E2,B),
  !,
  prod_loop(Ord2,B,E2-Ys,E1-[A|Xs]).
prod_loop(Ord1,_A,E1-_Xs,_E2-Ys):-
  flip(Ord1,_Ord2,X,Y,Pair),
  X in E1,member(Y,Ys),
  engine_yield(Pair),
  fail.
\end{code}

The predicate {\tt flip/5} builds a pair in the appropriate order
when the generators take turns being active in the recursive loop.
\begin{code} 
flip(1,2,X,Y,X-Y).
flip(2,1,X,Y,Y-X).
\end{code}
Here is the product of the natural numbers with themselves as an example:
\begin{codex}
?- nat(N),nat(M),prod(N,M,R),show(12,R).
[0-0,1-0,1-1,0-1,2-1,2-0,2-2,1-2,0-2,3-2,3-1,3-0]
\end{codex}

The singleton stream with a known constant (e.g., {\tt o}) is the neutral
element for the product, if we consider any element {\tt X} to be isomorphic to
{\tt o-X} and {\tt X-o}.  Moreover, the product is associative if we ignore the
association in the pair representation (e.g, {\tt 1-(2-3)} seen as equivalent to {\tt (1-2)-3})
and commutative if the order of the elements does not matter, when interpreting our streams as multisets.  Finally, under the same
assumptions, the product distributes over the sum.

\paragraph{Note:} Our {\tt lazy\_streams} package also provides the {\tt prod\_/3} stream product 
operation that avoids the engine creation and collection overhead,
coming from the use of the constructor {\tt eng/3} by using a \emph{Cantor
unpairing function} to split natural numbers generated by {\tt nat/1}
that are used to index dynamic arrays growing with each new element consumed
from the two input streams. By using the  $N \rightarrow N^k$ generalized Cantor
untupling function, implemented for instance in \cite{serpro}, one can obtain efficient
generator product operations for $k$ generators.

\subsection{An Embedded Language Interpreter}

With our sum and product operations ready, we can proceed with the design of
the embedded language, facilitating more complex forms of generator
compositions. The grammar of this embedded language of generator expressions $F$ is:
\begin{center}
\begin{tabular}{rcl@{\hspace{0.4cm}}l@{\hspace{1cm}}cl@{\hspace{0.4cm}}l}
  $F$    & ::=  & $F_1 + F_2$         & \it (sum)         & $\mid$ & $\{F\}$           & \it (setification)\\
         & $\mid$ & $F_1 \times F_2$  & \it (product)     & $\mid$ & $X^G$            & \it (engine) \\
         & $\mid$ & $N:M$             & \it (range)       & $\mid$ & $A$               & \it (constant)\\
         & $\mid$ & $[X|Xs] \mid []$  & \it (list)        & $\mid$ & $E$               & \it (stream)
\end{tabular}
\end{center}
The language comprises lists, engines, ranges, constants and existing streams, as well as their sums, products and
\emph{setification} (i.e., removing duplicates).
We have implemented it with a simple interpreter,
the predicate  {\tt eval\_stream(+GeneratorExpression, -Generator)},
which turns a generator expression into a ready-to-use generator that combines 
the effects of its components.
\begin{code}
eval_stream(E+F,S):- !,eval_stream(E,EE),eval_stream(F,EF),sum(EE,EF,S).
eval_stream(E*F,P):- !,eval_stream(E,EE),eval_stream(F,EF),prod(EE,EF,P).
eval_stream(E:F,R):- !,range(E,F,R).
eval_stream([],L):-!,list([],L).
eval_stream([X|Xs],L):-!,list([X|Xs],L).
eval_stream({E},SetGen):-!,eval_stream(E,F),setify(F,SetGen).
eval_stream(X^G,E):-!,eng(X,G,E).
eval_stream(A,C):-atomic(A),!,const(A,C).
eval_stream(E,E).
\end{code}
We already explained the auxiliary predicates used in the evaluator, except for
{\tt setify/2}. That predicate wraps a generator to ensure it
produces a set of answers, with duplicates removed, using the built-in {\tt distinct/2}\footnote{
\url{https://www.swi-prolog.org/pldoc/doc_for?object=distinct/2}
}.  
\begin{code}
setify(E,SE):-eng(X,distinct(X,X in E),SE).
\end{code}
This works for infinite generators (within the limits of available memory), in contrast
with sorting, which also removes duplicates, but assumes the list is finite.

Lastly, we define {\tt in\_/2} as a variant of \texttt{in/2} that takes a 
generator expression rather than a generator.
\begin{code}
:-op(800,xfx,(in_)).

X in_ GenExpr:-eval_stream(GenExpr,NewGen),X in NewGen.     
\end{code}
Here is a small example that combines many of the features described above:
\begin{codex}
?- X in_ ({[a,b,a]}+(1:3)*c).
X = a ; X = 1-c ; X = b ; X = 1-c ; X = 2-c; X = 1-c ; X = 2-c ....
\end{codex}

\section{Other Stream Generator Operations}\label{other}

This section describes the additional stream operations in our {\tt
lazy\_streams} package.

\subsection{Lazy Functional Programming Constructs}\label{sec:other:fp}

\paragraph{Map}

The {\tt map/3} predicate creates a generator that applies a binary predicate
to the subsequent elements in a given stream to produce a new stream.
\begin{code}
map(F,Gen,map_next(F,Gen)).

map_next(F,Gen,Y):-ask(Gen,X),call(F,X,Y).
\end{code}

Our package contains similarly defined {\tt map/N+1} generators that
apply a predicate with {\tt N arguments} to
{\tt N-1} stream generators.

\paragraph{Reduce}

The predicate {\tt reduce(+Closure,+Generator,+InitialVal, -ResultGenerator)}
creates a generator that reduces a finite generator's elements with the given closure, 
starting with an initial value. Its only element is the resulting single final value.
Similarly to Haskell's {\tt foldl} and {\tt foldr}, it can be used to
generically define arithmetic sums and products over a stream.

\begin{code}
reduce(F,InitVal,Gen, reduce_next(state(InitVal),F,Gen)).
\end{code}
It uses the predicate {\tt reduce\_next/4} 
that  applies closure {\tt F} to the state {\tt S}, 
while generator {\tt E} provides ``next'' elements.
\begin{code}
reduce_next(S,F,E,R):- \+ is_done(E),
  do((
    Y in E, arg(1,S,X),
    call(F,X,Y,Z),
    nb_setarg(1,S,Z)
  )),
  arg(1,S,R).

do(G):-call(G),fail;true.
\end{code}
Note that by working in $O(1)$ space, with destructive updates,
we can handle finite streams with an arbitrary number of elements.

\paragraph{Scan}

The predicate {\tt scan(+Closure, +Generator, +InitialVal, -ResultGenerator)} is
similar to \texttt{reduce/4} but also yields all intermediate results. Unlike the latter, this is
also meaningful for infinite streams.
\begin{code}
scan(F,InitVal,Gen,scan_next(state(InitVal),F,Gen)).

scan_next(S,F,Gen,R) :- arg(1,S,X),
  ask(Gen,Y),
  call(F,X,Y,R),
  nb_setarg(1,S,R).
\end{code}
For example, we can compute the stream of cumulative sums of the natural numbers.
\begin{codex}
?- nat(E), scan(plus,0,E,F), show(11,F).
[0, 1, 3, 6, 10, 15, 21, 28, 36, 45, 55].
\end{codex}

\subsection{Stream Wrappers for I/O and Stateful Prolog Features}

We can easily wrap file or socket readers as generators. This has the
advantage that details like opening a file, reading and
closing a stream stay hidden, as shown by the 
{\tt term\_reader/2} generator below.

\begin{code}
term_reader(File,next_term(Stream)):-open(File,read,Stream).

next_term(Stream,Term):-read(Stream,X),
  ( X\==end_of_file->Term=X
  ; close(Stream),fail
  ).
\end{code}

\section{Lazy Streams as Lazy Lists}\label{ll}

In addition to the abstract datatype representation for lazy streams where the
user interacts with them through a dedicated API, we also provide a more concrete
representation for lazy streams in the form of {\em lazy lists}. Lazy lists
look much like regular Prolog lists and, just like regular lists, they can be
inspected and deconstructed with unification. The difference with regular lists
is that lazy lists are not fully materialized from the start, but that, like
lazy streams, their elements are computed on demand and thus can conceptually
hold infinitely many elements. 

A key advantage over the abstract stream representation is that much of the
existing functionality for regular lists can be reused for lazy lists; a good
example are DCGs, which can also be used to parse lazy lists.

\subsection{Lazy List Representation}

Our lazy list representation is based on the lazy function technique of Casas
et al.~\cite{lazyCiao,casas2005functional}. The main idea is to delay the
evaluation of a computation until its result is actually needed. Initially, the
result is represented by a logic variable. When this variable is inspected
through unification~(the need), a coroutine mechanism (e.g., {\tt freeze/2} or
attributed variables~\cite{holz92}) is triggered to perform the computation and
deliver the actual value just in time for the inspection.

We apply this technique to compute a list incrementally as more
and more of it is needed. Thus a lazy \emph{list} is represented as a normal
Prolog list where the \emph{tail} is formed by an attributed variable. The following
code illustrates this approach on the lazy list of natural numbers.
\begin{code}
simple:lazy_nats(List):-simple:lazy_nats_from(0,List). 
simple:lazy_nats_from(N,List):-put_attr(List,simple,N).

simple:attr_unify_hook(N,Value):-succ(N,M),
  simple:lazy_nats_from(M,Tail),Value = [N|Tail].
\end{code}
A call to \texttt{simple:lazy\_nats/1} creates an attributed variable for the lazy list
that can be passed to a typical list traversal predicate.  Such a predicate
unifies this variable with either the empty list (\texttt{[]}) or a list cell
(\texttt{[Head|Tail]}), which triggers the attributed variable hook and
instantiates the variable to {\tt [N|Tail]} where {\tt Tail} is again an
attributed variable. When the list traversal predicate recurses on {\tt Tail},
this process is repeated.

\paragraph{Improvement}
There is a significant disadvantage to the above basic approach.  As already
observed by Casas et al., on backtracking, the lazily computed extension of the
list is lost and possibly recomputed again on the next forward computation. A
pathological case is that where the list traversal first tries to unify with
the empty list and then with the non-empty list. For such a predicate, every
element is computed twice, a first time when the unification with the empty
list fails, and a second time when the unification with the non-empty list
succeeds.

In addition to the recomputation overhead, this makes the implementation
unsuitable for fetching data from an external source---like a network socket---that cannot backtrack. It is possible to keep a
buffer to support re-fetching content from the socket but the amount of data we
need to buffer depends on the unknown non-determinism in the Prolog code that
processes the list and we cannot recover if the selected buffer size proves to
be too short.

We provide a solution for this problem by using \emph{non-backtrackable
assignment} in the form of SWI-Prolog' \texttt{nb\_setarg/3}, which assigns an
argument in a compound term and is not undone on backtracking. We illustrate this
idea on the lazy list of natural numbers:
\begin{code}
lazy_nats_from(N,L) :- put_attr(L,lazy_streams,state(N,_)).

attr_unify_hook(State,Value) :-State=state(N,Read),
  ( var(Read) -> succ(N,M),nats(M,Tail),
     nb_setarg(2,State,[N|Tail]),arg(2,State,Value)
  ;  Value = Read
  ).
\end{code}
Compared to the basic version, we use a compound state here, where a
the \texttt{Read} field is initially free 
and instantiated with the resulting list structure once the value has been computed. Because
of the use of {\tt nb\_setarg/3}, the information recorded in \texttt{Read}
survives backtracking.

With the above technique we have implemented the \texttt{pure\_input} library, which supports a lazy list view of files and sockets, 
as well as the generic
\texttt{lazy\_lists}
library. The general goal to create a lazy list is
\texttt{lazy\_list(:Next, +State0, -List)}. This executes
\texttt{call(Next, State0, State1, Head)} to produce the next element.

Lazy lists allow Prolog to handle infinite data streams in limited memory,
provided that garbage collection can reclaim the already processed part of the
list. This is possible if the user code does not keep a reference to the head
of the list. A particular pitfall here is nondeterminism: even when the current
branch no longer needs the head of the list, the runtime environment may have
to hold onto it for the sake of unexplored alternative branches.
Hence, non-determinism can only be mixed with lazy lists if every choicepoint
is resolved (i.e., no unexplored alternatives remain) after examining only a bounded number of additional elements. 
If this condition is met, 
the attributed variable trigger, which advances the stream, and garbage
collection, which reclaims the unused prefix of the list, together
ensure that the in-memory window of the stream is finite. 


%


\subsection{From Lazy Streams to Lazy Lists}

Now we show how to convert from lazy streams to lazy lists and back by
providing an isomorphism between both representations.
The former conversion is interesting because lazy lists may present a convenient, tangible 
representation. In contrast, the latter may be more convenient for defining new
generators, and avoids confusion with regular lists.
Indeed, although infinite lazy lists look like regular lists, they don't work
well with all regular list predicates. For instance,
\begin{codex}
?- lazy_nats(Ns),maplist(succ,Ns,Ps).
... loops forever ...
\end{codex}
The problem is that while the list is infinite and lazy, {\tt maplist/3} is eager
and only works on finite lists.
By exploiting the isomorphism between the two representations we can easily import the lazy {\tt map/3} from
streams to get a {\tt lazy\_maplist/3}.

\paragraph{Isomorphism} Two predicates witness the isomorphism between the
representations. The predicate {\tt
gen2lazy(+Generator,-LazyList)} turns a possibly infinite stream generator into
a lazy list by using the generator as the state on which the lazy list is
based, and using \texttt{ask/2} to advance that state (which is in fact already
handled by the generator), and produce a new element.
\begin{code} 
gen2lazy(Gen,Ls):-lazy_list(gen2lazy_forward,Gen,Ls).

gen2lazy_forward(E,E,X):-ask(E,X).
\end{code}
The opposite direction is even easier, as the {\tt list/2} generator also works
on lazy lists.
\begin{code}
lazy2gen(Xs, Gen):-list(Xs, Gen).
\end{code}

\paragraph{Iso-Functor}
We can easily transport not just the data representations but also the
operations acting on them. In category theory, this concept is formally known
as an {\em iso-functor}, a mapping that transports morphisms between objects
from one category to another and back. 

The predicate 
{\tt iso\_fun(+Operation, +SourceType, +TargetType, +Arg1, -Result)}
generically transports a predicate of the form {\tt F(+Arg1, -Arg2)} to a domain where
an operation can be performed and brings back the result.

\begin{code}
iso_fun(F,From,To,A,B):-call(From,A,X),call(F,X,Y),call(To,Y,B).
\end{code}
We have also defined similar code for predicates with other arities and modes.

\begin{codeh}
%
iso_fun(F,From,To,A,B,C):-
  call(From,A,X),
  call(From,B,Y),
  call(F,X,Y,Z),
  call(To,Z,C).

%
iso_fun_(F,From,To,A,B,C):- 
  call(From,A,X),
  call(F,X, Y,Z), 
  call(To,Y,B),
  call(To,Z,C).
\end{codeh}

This allows us to define lazy version of {\tt maplist}:
\begin{code}
lazy_maplist(F,LazyXs,LazyYs):-iso_fun(map(F),lazy2gen,gen2lazy,LazyXs,LazyYs).
\end{code}
where {\tt map/3} is the stream generator from Section~\ref{sec:other:fp}. Here
is an example of the result.
\begin{codex}
?- lazy_nats(Ns),lazy_maplist(succ,Ns,Ps),prefix([A,B,C],Ps).
Ns = [0,1,2|_20314], Ps = [1,2,3|_20332], A=1,B=2,C=3,...
\end{codex}

Inversely, an alternative {\tt sum\_/3} operation  can be implemented 
quite easily with lazy lists. Our {\tt lazy\_streams} package 
uses this technique to borrow it with help from the {\tt iso\_fun/6} predicate.
\begin{code}
sum_(E1,E2, E):-iso_fun(lazy_sum,gen2lazy,lazy2gen,E1,E2, E).

lazy_sum(Xs,Ys,Zs):-lazy_list(lazy_sum_next,Xs-Ys,Zs).
  
lazy_sum_next([X|Xs]-Ys,Ys-Xs,X).
lazy_sum_next(Xs-[Y|Ys],Ys-Xs,Y).
\end{code}

\section{Discussion}\label{disc}

The abstract sequence interface of the {\tt lazy\_streams} package and
the concrete list-based view provided by the {\tt lazy\_lists} library offer
similar services, but as they interoperate with help of {\tt iso\_fun} predicates,
one can choose the  implementation most suitable for a given algorithm.
For instance, one can access the $n$th element of a generator in $O(1)$ space.
Lazy lists might or might not need $O(n)$ for that, depending on possible garbage collection of their unused prefix.
With most stream generators no garbage collection is needed when working destructively in constant space, while their results can be exposed declaratively via the stream algebra.
On the other hand, lazy lists are reusable, while new generators 
must be created to revisit a sequence.
The Appendix in the extended version of this paper\footnote{
https://github.com/ptarau/AnswerStreamGenerators/blob/master/doc/techcom.pdf
}
also shows with benchmarks that stream generators are faster than lazy lists.

Some algorithms can be most easily expressed using first-class logic engines, but
avoiding engines when possible reduces memory footprint and can avoid term copying.
Thus, one might ask if a predicate like {\tt lazy\_findall} could be implemented
without using first class logic engines, e.g., it terms of  
attributed-variables, 
TOR \cite{tor}, delimited AND-continuations \cite{delim}.
This seems unlikely as  these techniques are all subject to backtracking and cannot
store state that survives it. First-class engines can be simulated~\cite{padl09inter}, but that is 
impractically slow. A more promising alternative is reflecting Prolog's backtracking mechanism by implementing OR-continuations at abstract machine level.

\section{Related work}\label{rel}

A relational view of stream processing and querying has been 
present in the database community 
\cite{Law11,Babcock02},
and 
within logic programming 
\cite{Beck15}, 
the latter with focus on
reasoning about streams.
Our lazy stream generators share with Python  \cite{pyref} the encapsulation of
a stream into a mechanism providing its elements on-demand. But, by contrast to
Python's generator operations (see library {\em itertools}), we ensure 
that everything scales up to work on infinite streams.
Adoption of a very similar mechanism by other widely
used languages validates the claims of 
enhanced expressiveness  that generators can bring.
The basic idea of using coroutining in the form of first-class logic engines
has been present in the BinProlog system  \cite{bp2011} as early as 1995 and
the attributed variables mechanism \cite{holz92} that we used to implement 
lazy lists, has been present even earlier, originally introduced to support constraint programming. However, putting it all together
in the form of a mature API is a 
 novel contribution of this paper,
 as well as the uniform view, encapsulating into an open-source library 
 our mix of declarative and procedural implementation techniques. 
 
\section{Conclusions}\label{conc}

We have described a unified approach to program with finite and infinite stream
generators that enhances Prolog with operations now prevalent in widely used
programming languages like Python, C\#, go, JavaScript, Ruby and Lua, while
also supporting lazy evaluation mechanisms comparable to those in non-strict
functional languages like Haskell. As a special instance, we have defined
generators based on first-class logic engines that can encapsulate both
AND-streams and OR-streams of answers.  Moreover, we have provided an embedded
interpreter for our generator algebra to enable declarative expression of
stream algorithms in the form of compact and elegant code.

In addition, we have provided a lazy list representation for our streams, which
interacts nicely with unification and typical Prolog list code. Our iso-functor
supports  transport of operations between lazy lists and generators, which
allows us to choose the simplest or most efficient implementation of stream
operations. In terms of impact, there are {\tt 83} github sites using the {\tt lazy\_lists} library, and
 Ogborne's analysis tools for the Enron e-mail
corpus\footnote{\url{https://github.com/Anniepoo/enron}} already make good use
of our {\tt pure\_input} library based on them. We plan to explore further applications and
expose our libraries to the wider community.


\subsection*{Acknowledgments}
This work has been partially supported by NSF grant \verb~1423324~
and by the Flemish Fund for Scientific Research (grant G0D1419N).

\bibliographystyle{eptcs}
\bibliography{theory,tarau,proglang,biblio,new}

\end{document}